\documentclass[
    reprint,
    amsmath,amssymb,
]{revtex4-2}

\usepackage{graphicx}
\usepackage{dcolumn}
\usepackage[dvipsnames]{xcolor}
\usepackage{bm}
\usepackage[version=4]{mhchem}
\usepackage{ragged2e}
\usepackage{hyperref}
\usepackage{url}
\usepackage{upgreek}
\usepackage[T1]{fontenc}
\usepackage[utf8]{inputenc}

\hypersetup{
    colorlinks=true,
    linkcolor=blue,
    filecolor=magenta,      
    urlcolor=cyan,
    pdftitle={Overleaf Example},
    pdfpagemode=FullScreen,
    }
    
\begin{document}
\preprint{APS/123-QED}

\title{Mapping the strong-to-weak coupling crossover in polymer-film microcavity lasers}

\author{Denis A. Sannikov$^{1,\bullet}$}
\author{Nailya M. Urazova$^{1,\bullet}$}
\author{Maksim D. Kolker$^{1,\bullet}$}
\author{Grigorij D. Ivanov$^{1}$}
\author{Anton D. Putintsev$^{1}$}
\author{Aleksandr V. Averchenko$^{1}$}
\author{Daria Khmelevskaia$^{1}$}
\author{Liliya T. Sahharova$^{2}$}
\author{Nikita S. Shlapakov$^{2}$}
\author{Ioannis Paschos$^{3}$}
\author{Pavlos G. Savvidis$^{3}$}
\author{Valentine P. Ananikov$^{2}$}
\author{Pavlos G. Lagoudakis$^{1}$}
\thanks{Contact author: p.lagoudakis@skoltech.ru\\$^{\bullet}$ D.S., N.U., and M.K. contributed equally to this work.}

\affiliation{$^1$Hybrid Photonics Laboratory, Skolkovo Institute of Science and Technology Territory of Innovation Center Skolkovo Bolshoy Boulevard 30, building 1, Moscow 121205, Russia}
\affiliation{$^2$Zelinsky Institute of Organic Chemistry, Russian Academy of Sciences, Leninsky Prospect 47, 119991 Moscow, Russia}
\affiliation{$^3$Westlake University, Zhe Jiang Sheng, Hang Zhou Shi, Xi Hu Qu, 310024, China}

\date{\today}

\begin{abstract}
Organic semiconductors are particularly attractive for polaritonics due to their large exciton binding energies and oscillator strengths. Among them, the ladder-type conjugated polymer poly(paraphenylene) is distinguished by its rigid backbone, narrow exciton linewidth, high photoluminescence quantum yield, and enhanced photostability, making it an excellent candidate for organic polariton devices. While polariton lasing has been reported in various organic systems, systematic studies of the transition from polariton lasing to conventional photon lasing within a single, well-controlled material platform remain limited. Here, we present planar organic microcavities incorporating MeLPPP as the active medium, in which continuous tuning of the effective cavity length within a single device enables us to map the strong-to-weak coupling transition across five distinct cavity-mode orders. We demonstrate an approximately eighteen-fold increase in the lasing threshold when crossing from polariton to photon lasing. We further establish a quantitative framework in which the spectral dependence of the threshold governs a universal V-shaped blueshift of the emission energy across both coupling regimes. Finally, we show that vibron-mediated exciton relaxation, previously identified in the strong-coupling limit, persists across the crossover: lasing-threshold minima track the vibron resonances throughout the coupling transition.
\end{abstract}
                              
\maketitle

\section{Introduction} 

Strong coupling between confined photonic modes and semiconductor excitonic transitions in optical microcavities gives rise to hybrid light–matter quasiparticles known as exciton–polaritons \citep{Kavokin2007,Agranovich2003}. Organic semiconductors are particularly well suited for realizing room-temperature polariton phenomena \citep{Jiang2021,Feist2017,Lerario2017} owing to the large binding energies and oscillator strengths of Frenkel excitons. To date, polariton condensates have been demonstrated in diverse organic systems, such as organic crystals \citep{Wang2025}, oligofluorenes, fluorescent proteins \citep{Satapathy2022}, molecular dyes \citep{Cookson2017,Sannikov2019,Putintsev2020}, and conjugated polymers \citep{Wei2019}.

\begin{figure*}[t!]
    \centering
    \includegraphics[width=0.9\textwidth]{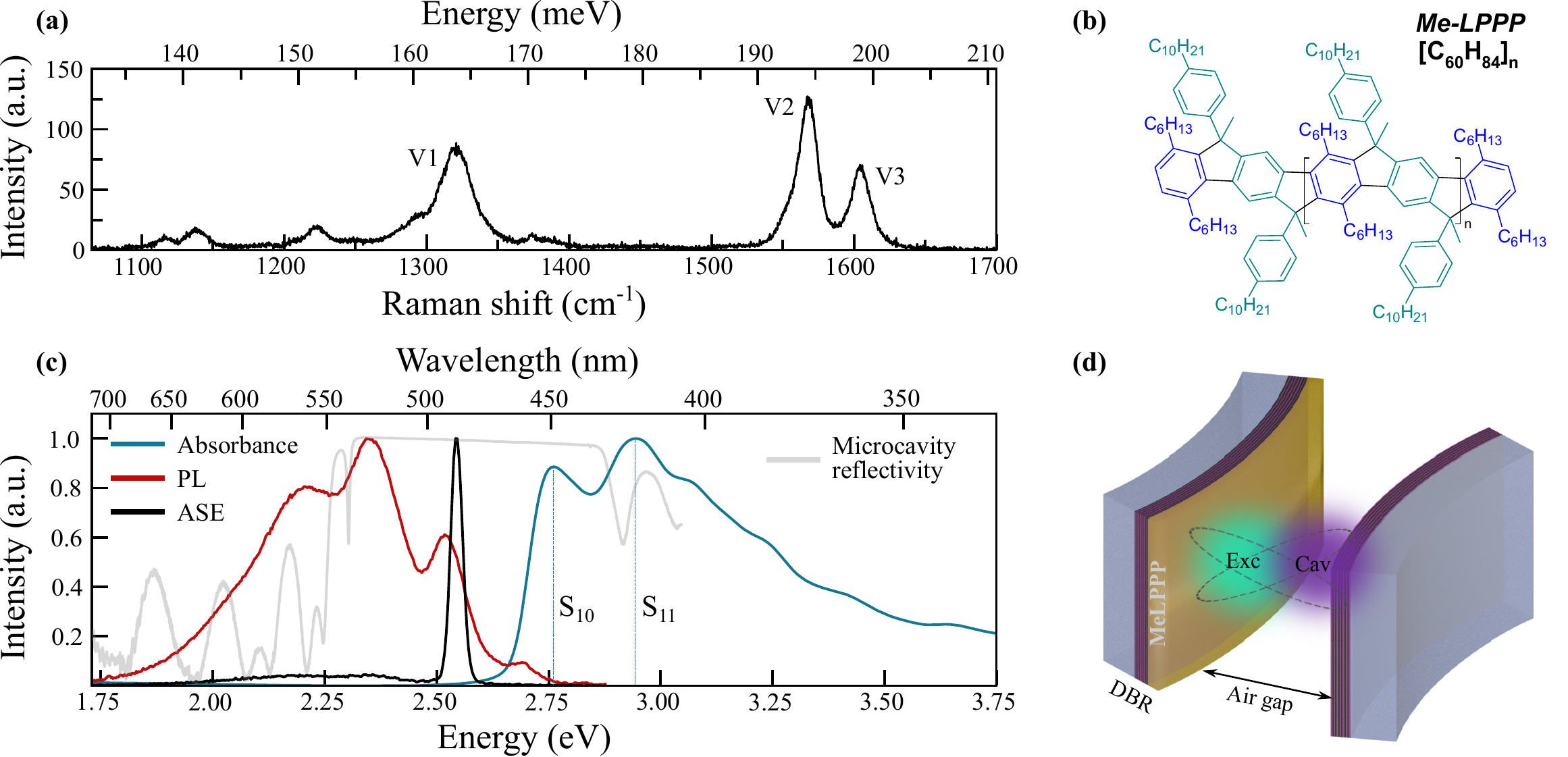}
    \caption{\justifying Material characterization. \textbf{a} Raman spectrum of methyl-substituted ladder-type polymer poly(paraphenylene) (MeLPPP) reference, 180-nm, non-cavity thin film with three highlighted molecular vibrational modes V1, V2, and V3. \textbf{b} Chemical structure of the polymer chain. \textbf{c} Normalized absorption (blue line) with two highlighted excitonic transitions $S_{10}$ and $S_{11}$, fluorescence (red line), amplified spontaneous emission 487.25 nm/2.545 eV (black line) spectra of the corresponding film. Gray line shows the microcavity reflectivity spectrum. \textbf{d} Schematic structure of the asymmetric open microcavity: a MeLPPP polymer film is spin-coated onto the bottom DBR (\ce{Ta2O5}/\ce{SiO2}, \ce{Ta2O5}-terminated), and a second DBR is positioned above the polymer layer with a variable air gap, forming an asymmetric open-cavity geometry whose effective cavity length is tuned by lateral translation of the sample.}
    \label{Fig1}
\end{figure*}

\begin{figure*}[t!]
    \centering
    \includegraphics[scale=0.42]{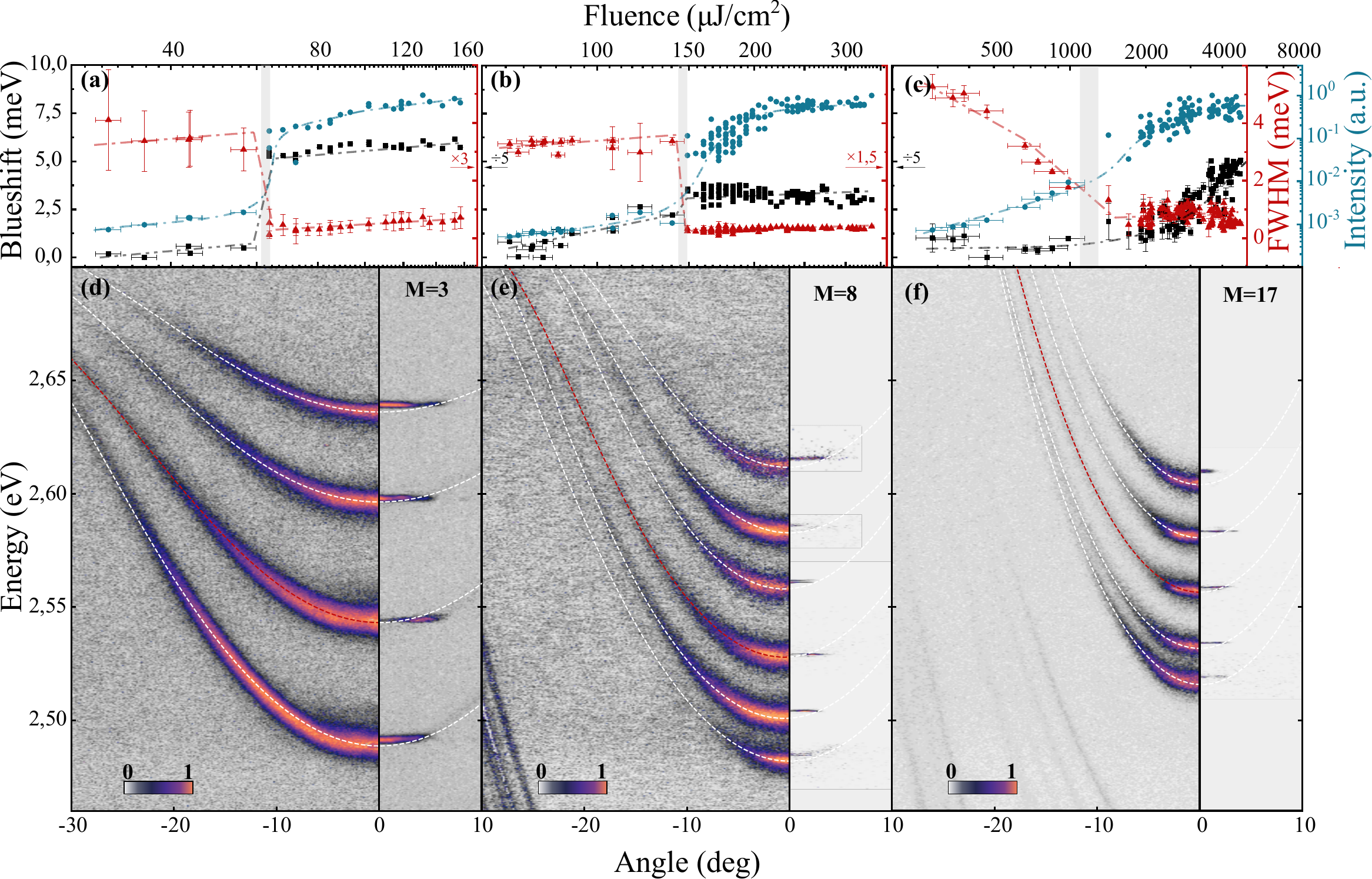}
    \vspace*{-0.25cm} 
    \caption{\justifying \label{Fig2} Hallmarks of polariton and photon lasing. \textbf{a}, The polariton PL intensity (blue circles), linewidth (red triangles), and energy shift (black squares) as a function of pumping fluence in a strongly coupled regime (order $M=3$, effective cavity length $L_{\rm eff}=731\ \mathrm{nm}$) with the polariton lasing threshold indicated with the vertical gray stripe, $P^{\rm pol}_{\mathrm{th}}=62.6\ \mu\mathrm{J}\,\mathrm{cm}^{-2}$.
    \textbf{b}, Same as \textbf{a} for the intermediate coupling regime (order $M=8$, $L_{\rm eff}=1962\ \mathrm{nm}$) with the lasing threshold $P_{\mathrm{th}}=145\ \mu\mathrm{J}\,\mathrm{cm}^{-2}$.
    \textbf{c}, Same as \textbf{a} for the weakly coupled regime (order $M=17$, $L_{\rm eff}=4122\ \mathrm{nm}$) with the photon lasing threshold $P^{\rm ph}_{\mathrm{th}}=1100\ \mu\mathrm{J}\,\mathrm{cm}^{-2}$. In all three panels, a step-like blueshift of the emission energy is observed at threshold. Dashed curves in panels \textbf{a}--\textbf{c} are guides to the eye. Energy shift values are multiplied by a factor of 5 (\textbf{b, c}). Linewidth values are divided by a factor of 3 (\textbf{a}), 1.5 (\textbf{b}).
    \textbf{d}--\textbf{f}, Dispersion images of PL below (left sub-panels) and above (right sub-panels) the lasing threshold for $M=3$ (\textbf{d}), $M=8$ (\textbf{e}), and $M=17$ (\textbf{f}), each shown at several different cavity lengths corresponding to different exciton--photon detuning conditions. White dashed lines in panels \textbf{d--f} represent the best-fit results of a coupled-oscillators model. The red dashed lines in panels \textbf{d-f} indicate the modes whose lasing hallmarks are plotted in the corresponding panels \textbf{a--c} above. The colour scale is normalised to the maximum intensity of the above-threshold dispersion images.}
    \vspace*{-0.5cm}    
\end{figure*}

Among these candidates, conjugated polymers have attracted considerable interest for both fundamental polaritonics and optoelectronic device applications such as light-emitting diodes \citep{Baigent1994}, polymer lasers \citep{Wenger2010}, photovoltaic cells \citep{Granstrm1998}, field-effect transistors \citep{Nikolka2016}, and batteries \citep{Xie2017,Mike2013}. They combine the electronic functionality of semiconductors with the mechanical flexibility and solution processability of plastics, enabling the fabrication of large-area, optical-quality thin films at low cost. The $\pi$-electron delocalization over chromophores in conjugated polymers enhances their transition dipole moment, giving rise to large Rabi splitting values (>500 meV) under strong coupling conditions even in simple planar microcavities \citep{Roux2022}.

Within this family, the methyl-substituted ladder-type polymer poly(paraphenylene) (MeLPPP) is particularly notable. Its rigid, rod-like backbone, stabilized by methylene bridges, produces narrow exciton linewidth with distinct vibronic structure, a small Stokes shift, and a high PL quantum yield up to 30\% \citep{Stampfl1995,Leising1995}, along with the enhanced photostability \citep{Kranzelbinder1997}. Microcavities incorporating MeLPPP have enabled the realization of exciton–polariton condensates \citep{Plumhof2013} and advanced polariton-based functionalities, such as tunable lasing \citep{Urbonas2016,Scafirimuto2017}, multimode lasing \citep{Urbonas2024}, ultrafast all-optical transistors \citep{Zasedatelev2019,Tassan2024}, cascadable logic gates \citep{Sannikov2024} leveraging sub-THz temporal dynamics \citep{Misko2025} and high degree of second-order coherence \citep{Putintsev2024} of polariton condensates, and quantum simulations in engineered photonic lattices \citep{Scafirimuto2021,Georgakilas2025}.

The transition between conventional photon lasing, based on population inversion, and polariton lasing, driven by bosonic stimulation of exciton–polariton scattering, has been studied in organic systems by different methods including changing cavity length \citep{Dietrich2016}, increasing molecular density in active layer (superabsorption) \citep{Quach2022}, and exploiting polarization-dependent birefringence in aligned perylene diimide films \citep{Herrmann2024}. Nevertheless, no results have been reported for a continuous range of cavity lengths within a single, photostable material platform with a well-resolved vibronic structure. Understanding the crossover between these regimes is essential for distinguishing polariton-specific signatures from conventional lasing and for optimizing device design.

Vibron-assisted relaxation has been established as an important mechanism in strongly coupled organic microcavities \cite{Coles2011, Somaschi2011}, and the deliberate exploitation of vibronic resonances under resonant pumping has been used to reduce condensation thresholds by an order of magnitude \cite{Zasedatelev2019}. Whether this channel persists within the strong-coupling regime, and how the threshold modulation evolves as the system is tuned towards the weak-coupling regime, has not, however, been systematically addressed.

In this work, we present a straightforward fabrication procedure for planar, strongly coupled organic microcavities incorporating the conjugated ladder-type polymer MeLPPP as an active medium capable of supporting polariton lasing. By systematically varying the effective cavity length within a single device, we track the transition from strong to weak coupling through the polariton condensation threshold, emission linewidth, and energy shift as functions of non-resonant optical pump fluence. The single-device mapping, spanning five cavity mode orders that straddle the Savona criterion $\hbar\Omega_R > \Gamma_{\rm exc}+\Gamma_{\rm cav}$ \cite{SAVONA1995}, provides the continuous detuning trajectory. On this trajectory we derive and test a quantitative link between the spectral dependence of the lasing threshold and the spectral dependence of the nonlinear energy shift at threshold. Building on the saturation-driven blueshift mechanism of Ref.~\citep{Yagafarov2020}, in which the emission energy is pulled by Pauli-blocking-induced saturation of the molecular transitions, we show that the magnitude of the shift is set by the lasing threshold itself, which is in turn governed by the gain spectrum of the active medium. Since the gain spectrum is centred on the amplified spontaneous emission (ASE) peak, the predicted blueshift follows a parabolic (V-shaped) dependence as a function of the spectral offset $\Delta E = E_{\rm ASE}-E_{k_0}$ between the ASE maximum and the lasing mode. The framework also predicts that the V-shape acquires a non-zero baseline value at $\Delta E=0$ in the strong-coupling regime, arising from a quenching of the Rabi splitting under saturation, and that this baseline vanishes once strong coupling is lost, providing, in principle, a coupling-regime-dependent diagnostic in addition to the universal V-shape itself. Within the same dataset, the vibron-mediated-relaxation picture of Refs.~\citep{Zasedatelev2019,Coles2011,Somaschi2011} is tested within the strong-coupling regime: we find that the lasing-threshold minima track the V$_2$ and V$_3$ vibron resonances for the strongly coupled mode orders and that this modulation smoothly weakens as the coupling strength is reduced towards the weak-coupling limit. Throughout, we employ non-resonant, ultrafast (250 fs, 400 nm) excitation, which populates the molecular density uniformly and therefore interrogates the coupling-regime dependence without the spectral selectivity inherent in resonant vibronic pumping. The combined result is an approximately eighteen-fold increase in the lasing threshold between the polariton and photon lasing regimes at their respective optimal exciton-photon detunings relative to the material gain maximum, and a decomposition of the energy-shift data that distinguishes universal gain-driven behaviour from polariton-specific signatures.

\section{Experimental results} 

We synthesize Me-LPPP precipitate following published procedures \citep{Ammenhauser2020,Zhai2021,Bonn2016,Scherf1991,noauthor1984,Scherf1992}, and characterize the resulting material by measuring the Raman spectrum of a 180-nm thin film, spin-coated onto \ce{SiO2}/\ce{Si} substrate, see Fig.\ref{Fig1}a, with the chemical structure of the polymer depicted in Fig.\ref{Fig1}b. We observe nearly identical spectrum as reported previously \citep{Vaughan2008,Zasedatelev2019}, with the slightly varying relative intensities of V1-, V2-, and V3 vibron bands, for details on the synthesis procedure see Section 1 in SI. The normalized absorption spectrum of the MeLPPP thin film is shown in Fig.\ref{Fig1}c depicted with blue solid line. Two well-resolved peaks in the absorption spectrum are assigned to optical transitions from the ground singlet state \ce{S0} to sub-levels of the first excited singlet state \ce{S1}: the purely electronic (0-0) transition at 449 nm (2.76 eV) and the first vibronic (0-1) sideband at 421 nm (2.94 eV). Their Stokes-shifted counterparts in the fluorescence spectrum shown with red solid line in Fig.\ref{Fig1}c are observed at 463 nm (2.68 eV) and 493 nm (2.51 eV), respectively. The material's ability to manifest optical gain was characterized by measuring ASE along a stripe excitation profile, see Section 2 in SI for setup schematics. ASE spectrum of a 180-nm thin film is shown with black solid line in Fig.\ref{Fig1}c. The gain peak, identified by a nonlinear rise in emission intensity at a threshold incident pump fluence of 4.9 $\mu$J cm$^{-2}$, is centred at 487.25 nm (2.545 eV) and approximately coincides with (0-1) vibrational transition, suggesting a four-level lasing scheme \citep{Cookson2017}. 

\begin{figure}[t!]
    \centering
    \includegraphics[width=0.45\textwidth]{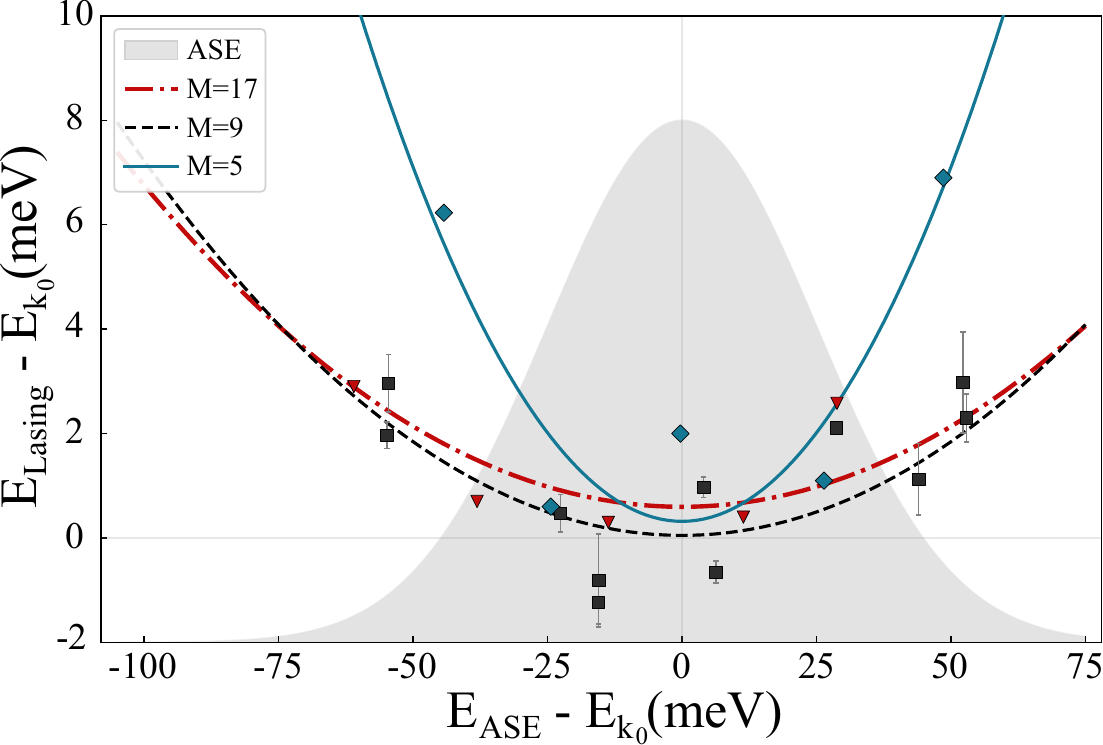}
    \caption{\justifying Energy shifts of lasing emission. Energy shift of the nonlinear PL emission at $\approx 2\times P_{\rm th}$ with respect to the linear, below-$P_{\rm th}$ emission from $k_{0}\equiv k_{||} = 0\pm5~^\circ$, for cavity orders $M=5$ (yellow diamonds, strong coupling), $M=9$ (black squares, intermediate coupling), and $M=17$ (green triangles, weak coupling) as a function of energy offset between the ASE maximum, $E_{\rm ASE}=2.545~\rm eV$, and the ground-state energy of the dispersion, $E_{k_{0}}$. Solid and dashed curves show the best fits by the parabolic model $E_{\rm lasing} - E_{k_0} = A_{\rm sat} \cdot (\Delta E / w)^2 + E_0$. The grey shaded area shows the normalised ASE spectrum of a $180$-nm thin MeLPPP film.}
    \label{Fig3}
    \centering
\end{figure}

We fabricate microcavities by spin-coating a 180-nm-thick MeLPPP film onto a high-reflectivity distributed Bragg reflector (DBR), consisting of 12.5 pairs of alternating \ce{Ta2O5}($n\approx2.14$)/\ce{SiO2}($n\approx1.46$) quarter-wavelength thick layers, with the high-index \ce{Ta2O5} layer terminating at the polymer interface to maximize field overlap with the active medium (see Methods section for details).
A second DBR is then positioned above the polymer layer using a point-squeeze mounting apparatus \citep{Kolker2024}, creating an asymmetric open-cavity geometry with an air gap between the polymer surface and the top mirror. This configuration yields a $Q-$factor of $\approx2000$ within the stop-band center. Numerical calculations confirm that one of the electric-field antinodes is located at the polymer layer for the cavity modes of interest (for details see Section 3 in SI). A one-time compression procedure then induces the curvature in the top and bottom DBRs, forming a convex-shaped microcavity with an optical cavity thickness increasing radially from a minimum at the compression point. The effective cavity length is therefore tuned by translating the sample laterally, which displaces the pump spot to regions of different thickness relative to the compression point. The sample geometry is illustrated schematically in Fig.\ref{Fig1}d.

\begin{figure*}[t!]
    \centering
    \includegraphics[width=1\textwidth]{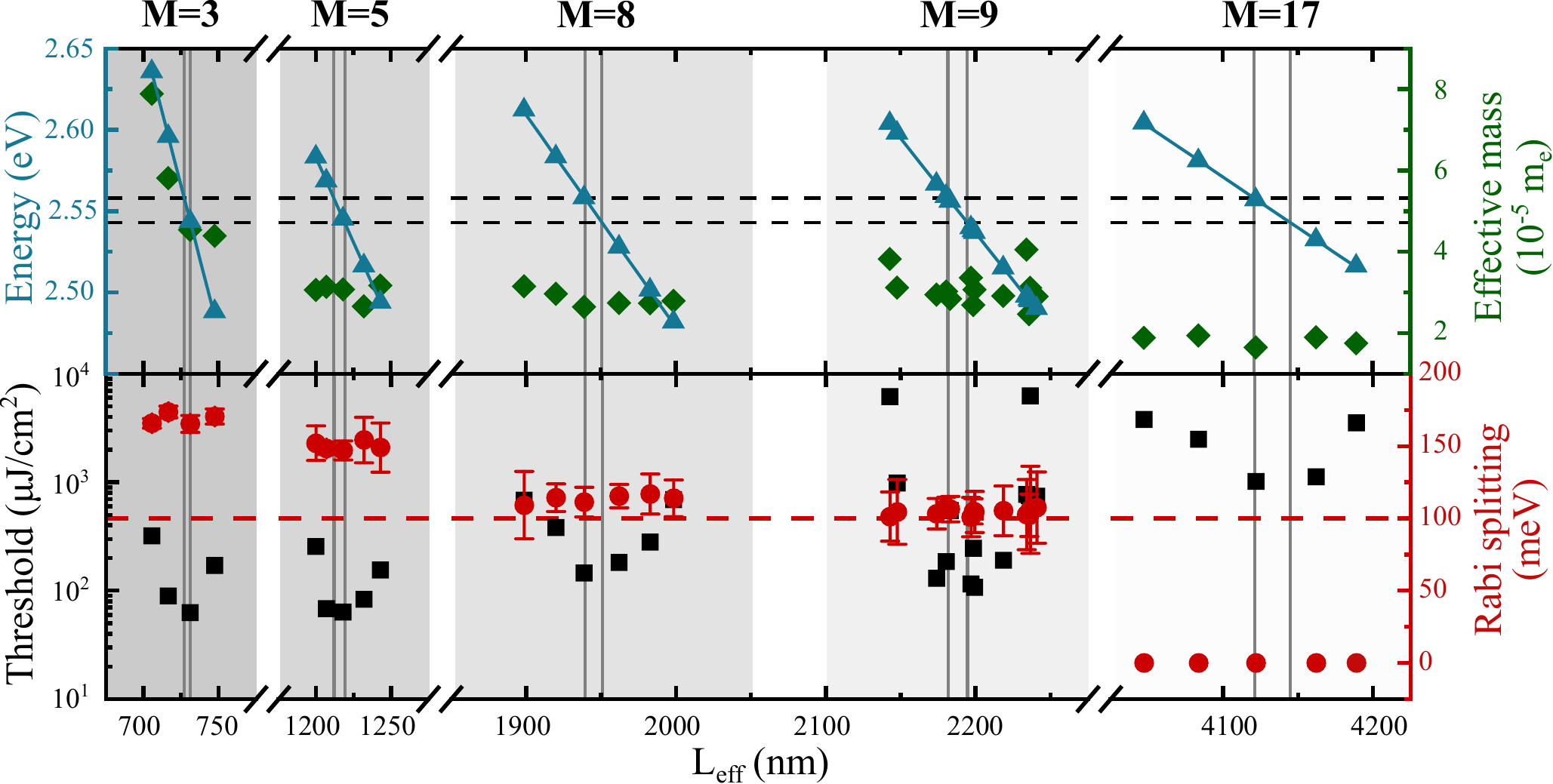}
    \caption{\justifying Cavity length tuning dependencies. Lasing threshold fluence (black squares markers, left axis of the bottom panel), Rabi splitting constant $\hbar\Omega_R$ (red circle markers, right axis of the bottom panel), emission energy from $k_{\parallel}=0$ of either the LPB or the cavity photon mode (blue triangle markers, left axis of the top panel), and polariton effective mass (green diamond markers, right axis of the top panel) are plotted as functions of the effective cavity length $L_{\mathrm{eff}}$ for five mode orders, $M=3,\ 5,\ 8,\ 9,\ 17$. According to the Savona criterion $\hbar\Omega_R > \Gamma_{\rm exc} + \Gamma_{\rm cav} = 99.9\pm 2.4$ meV (red horizontal dashed line in the bottom panel) \citep{SAVONA1995}, modes $M=3$ and $M=5$ ($\hbar\Omega_R \approx 140$--$170$ meV) correspond to the strong-coupling regime, modes $M=8$ and $M=9$ ($\hbar\Omega_R \approx 100$--$110$ meV) lie in the immediate vicinity of the criterion and are classified as intermediate, while $M=17$ ($\hbar\Omega_R \rightarrow 0$) corresponds to the weak-coupling regime. The effective mass decreases gradually from $\sim 8\times 10^{-5}\,m_e$ at small cavity lengths to $\sim 2\times 10^{-5}\,m_e$ at large cavity lengths, providing a model-independent signature of the strong-to-weak coupling transition. Horizontal-dashed (upper panel) and vertical-solid (bottom panel) lines indicate spectral positions of V2 and V3 vibron resonances with respect to the $S_{10}$ exciton.}
    \label{Fig4}
    \centering
\end{figure*}

To distinguish the strong- and weak-coupling regimes quantitatively, we adopt the standard criterion of Savona \textit{et al.} \citep{SAVONA1995}: strong coupling requires the Rabi splitting $\hbar\Omega_R$ to exceed the sum of the exciton and cavity linewidths, $\hbar\Omega_R > \Gamma_{\rm exc} + \Gamma_{\rm cav}$. From the absorption spectrum of a 180-nm MeLPPP reference film we extract $\Gamma_{\rm exc} = 92.0 \pm 2.4$~meV (FWHM), and from the below-threshold cavity dispersions we extract $\Gamma_{\rm cav} = 7.86 \pm 0.28$~meV (FWHM), yielding a criterion threshold of $\Gamma_{\rm exc} + \Gamma_{\rm cav} = 99.9 \pm 2.4$~meV. In what follows we characterise the lasing behaviour at three representative cavity-mode orders -- $M=3$ (strong coupling, $\hbar\Omega_R \approx 170$~meV > the criterion), $M=8$ (in the immediate vicinity of the Savona criterion, classified as an intermediate regime), and $M=17$ (weak coupling, $\hbar\Omega_R \to 0$).

Under normal-incidence, 250-fs, non-resonant optical excitation spectrally tuned at 400 nm, the cavity is optically pumped from the polymer-DBR side (bottom) to maximize absorption in the active layer, and photoluminescence is collected in transmission through the air-gap side (top). We perform either a multi-pulse integrated imaging of PL emission for the below-lasing-threshold regime or a single-shot PL imaging of the above-lasing-threshold emission in k-space as a function of excitation density recorded in transmission configuration (see Methods) for a set of cavity lengths. This asymmetric excitation-collection geometry has been shown to influence mode characteristics in open cavities \cite{Kedziora2024}. Across all measurements, we tune the effective cavity length in a wide range from 700 nm to 4.2 $\mu$m. For length values between 700 and 1250 nm, that correspond to the strongly coupled regime, we observe a non-linear increase in the PL intensity integrated over $\pm5~^\circ$ around normal incidence, concomitant narrowing of the emission full-width at half maximum (FWHM) line width, and the step-like blueshift of the emission energy, as evidenced by blue, red, and black axes in Fig.\ref{Fig2}a, respectively. This set of data is acquired for 731-nm effective cavity thickness and the third-order cavity photon mode ($M=3$) and is chosen to demonstrate an indicative behaviour of the studied regime. These three characteristics constitute the key signatures of polariton lasing \cite{Putintsev2023,Yagafarov2020}, providing clear evidence for its occurrence.  

Figure \ref{Fig2}d shows the dispersion images of the below (left sub-panel) and above (right sub-panel) the lasing threshold PL for four distinct detuning conditions $\Delta=$ -58, -107, -174, and -235 meV (M=3). Above the polariton lasing threshold $P_{\rm th}^{\rm pol}$, the k-space PL distribution collapses towards $k_{||}=0 \pm5~^\circ$, indicating the onset of coherent macroscopic occupation at the ground, lowest energy state. Together with the gradual broadening of the polariton lasing linewidth above the threshold, these hallmark behaviours evidence the presence of the strong light-matter coupling enabling the emergence of new polariton eigenstates inside the system. To quantify the interaction strength of the system, namely the Rabi-splitting energy constant, we apply a coupled-oscillators model with damping and fit the experimentally observed PL dispersion images, as indicated by white dashed lines of lower and upper polariton branches (LPB and UPB) in Fig.\ref{Fig2}d. The Hamiltonian of the system reads as follows:

\begin{equation}
    \hat{H} = \begin{pmatrix}
    E_{{\rm cav}|\rm\textbf{k}}-i\gamma_{cav} & V\\
    V & E_{exc}-i\gamma_{exc}
    \end{pmatrix}
\end{equation}

\noindent Here, $E_{\rm cav|\textbf{k}}$ is the eigenenergy of the cavity photon mode, $E_{exc}$ is the $S_{10}$ exciton energy, $\gamma_{cav} = \Gamma_{cav}/2$, $\gamma_{exc} = \Gamma_{exc}/2$, and $V$ is the interaction strength of the photon-exciton coupling for the $S_{00}\rightarrow S_{10}$ optical transition. Rabi splitting $\hbar\Omega_R$ is related to interaction strength $V$ by $\hbar\Omega_R=\sqrt{4V^2 - (\gamma_{cav}-\gamma_{exc})^2}$. In our case $|\gamma_{cav}-\gamma_{exc}| = 42$ meV and the square root that defines Rabi splitting value remains purely real for all analysed dispersions. We restrict the coupled-oscillator description to the $S_{10}$ exciton because the $S_{11}$ transition at $2.94$ eV falls outside the high-reflectivity stop-band of our DBRs, where the cavity $Q$-factor drops from $\approx 2000$ at the stop-band center to $\approx 70$ (see the reduced reflectivity of the cavity near the $S_{11}$ line of the absorption spectrum in Fig.\ref{Fig1}c); the cavity finesse is therefore insufficient for $S_{11}$ to participate in the observed polariton physics, and coupling to $S_{11}$ is effectively absent within the measurement range of this work. 

We now turn to the intermediate and weak-coupling regimes announced at the outset. At the intermediate effective cavity length $L_{\rm eff}=1962$~nm ($M=8$), the pump-fluence dependence and dispersion images are shown in Fig.\ref{Fig2}b and Fig.\ref{Fig2}e, respectively. Compared with the strongly coupled $M=3$ case, the lasing threshold rises by roughly a factor of four, the step-like blueshift at threshold persists but is reduced to less than half of its $M=3$ magnitude, and the dispersion deflection, though still identifiable in the coupled-oscillator fit (white dashed lines in Fig.\ref{Fig2}e), is markedly weaker. These intermediate-regime signatures are consistent with the Savona criterion placing $M=8$ in the immediate vicinity of the strong-coupling boundary rather than firmly inside it. At the largest cavity length investigated ($L_{\rm eff}=4122$~nm, $M=17$, Fig.\ref{Fig2}c,f), the lasing threshold increases by more than an order of magnitude relative to $M=3$, and the dispersion acquires a bare-cavity parabolic shape without visible anticrossing, indicating that the system has crossed into the weak-coupling regime where lasing occurs via conventional population inversion rather than bosonic stimulation.

Extending the Savona-criterion analysis to all five measured mode orders (Fig.\ref{Fig4}), $M=3$ and $M=5$ exhibit $\hbar\Omega_R=140$--$170$ meV and clearly satisfy the strong-coupling criterion; $M=8$ and $M=9$ yield $\hbar\Omega_R=100$--$110$ meV, in the immediate vicinity of the threshold, and are therefore classified as an intermediate regime; and $M=17$ shows $\hbar\Omega_R$ approaching zero together with a narrow, nearly-parabolic dispersion, consistent with the weak-coupling regime. As an independent indicator of the same crossover, we extract the effective mass of each measured dispersion (Fig.\ref{Fig4}, top panel), which gradually decreases from $\sim 8\times 10^{-5}\,m_e$ ($m_e$ is an electron mass) at small cavity lengths to $\sim 2\times 10^{-5}\,m_e$ at large cavity lengths, being a signature of the increasing photonic character of the lower branch as the coupling weakens.

A natural question is whether the nonlinear energy shift at threshold carries coupling-regime-specific information beyond what the lasing thresholds themselves reveal. Following the framework of Yagafarov \textit{et al.} \citep{Yagafarov2020}, the saturation-induced blueshift in organic microcavities is linear in the saturation parameter $\xi$, which under uniform non-resonant pumping is proportional to the threshold fluence $P_{\rm th}$. Because $P_{\rm th} \propto 1/\sigma_{\rm gain}(E_{k_0})$ and the gain cross-section $\sigma_{\rm gain}(E)$ is smooth with a maximum at $E_{\rm ASE}$, a second-order Taylor expansion yields a parabolic threshold dependence: $P_{\rm th} \propto [1 + (\Delta E / w)^2]$, where $\Delta E = E_{\rm ASE} - E_{k_0}$ and $w$ is a characteristic gain width, see Section 4 in the SI for a detailed analysis. This framework therefore predicts that the total energy shift should follow
\noindent 
\begin{equation}
    E_{\rm lasing} - E_{k_0} = A_{\rm sat} \cdot (\Delta E / w)^2 + A \cdot \Delta E + E_0
\end{equation}
with a symmetric, parabolic V-shape in $\Delta E$, an antisymmetric component $A\cdot\Delta E$ arising from any gain-induced frequency pulling, and a baseline $E_0$ set by coupling-regime-specific contributions (notably Rabi quenching, which should contribute in the strong-coupling limit and vanish in the weak-coupling limit). An independent estimate of the gain-pulling contribution yields less than $1$~meV across the entire spectral range, owing to the narrow cavity linewidths ($\Delta\omega_{\rm c}\sim1.5$--$10$~meV) relative to the broad ASE bandwidth ($\sim40$~meV); we therefore expect $A \approx 0$ and a symmetric V-shape across all coupling regimes.

Figure \ref{Fig3} tests this prediction directly, showing the measured energy shifts at $\approx 2\times P_{\rm th}$ relative to the linear (below-$P_{\rm th}$) emission at $k_{||}=0\pm5~^\circ$ for three representative cavity orders: $M=5$ (strong coupling, yellow diamonds), $M=9$ (intermediate, black squares), and $M=17$ (weak coupling, green triangles). The data follow the predicted parabolic V-shape across all three regimes, with RMS residuals of $0.8$--$1.1$~meV and no detectable antisymmetric component (consistent with $A=0$ within experimental precision). The baseline $E_0$, which the saturation framework of Ref.~\citep{Yagafarov2020} associates with Rabi-splitting quenching and which should therefore grow with the coupling strength in the strong-coupling regime and vanish in the weak-coupling limit, is extracted from the fit for each cavity order: $M=5$ (strong coupling, $\hbar\Omega_0 \approx 140$~meV) yields $E_0 = +0.3$~meV, while $M=9$ and $M=17$ yield near-zero baselines. The ordering of these values is consistent with the theoretical expectation, but the fitted $E_0$ for $M=5$ is comparable to the RMS residual of the fit itself, so within the present experimental precision the data do not allow a statistically meaningful verification of the baseline as a coupling-regime diagnostic. Disentangling this contribution from experimental scatter would require either measurements at deeper strong coupling (where a larger Rabi splitting should amplify the Rabi-quenching contribution) or a reduction in the per-point energy-shift uncertainty; both are natural directions for future work. The universality of the V-shape itself across all coupling regimes nevertheless confirms, in agreement with Ref.~\citep{Yagafarov2020}, that the energy-shift magnitude is governed primarily by the threshold-driven saturation parameter and does not by itself distinguish polariton from photon lasing.

Having established the three representative regimes, we now examine the full detuning dependence across all five measured mode orders (Fig.\ref{Fig4}). In the strong-coupling regime ($M=3$ and $M=5$), the interaction strength $\hbar\Omega_R$ for a specific cavity mode monotonically decreases with increasing detuning, as theoretically predicted (Fig.\ref{Fig4}, bottom panel).

The lowest threshold fluence for polariton lasing is achieved when the exciton-to-polariton transition matches the energy of the most pronounced vibronic resonances, V2 and V3, see vertical and horizontal dashed lines in Fig.\ref{Fig4}. The defining role of interactions between Frenkel excitons, polaritons, and molecular vibrational modes on the overall dynamics of the light-matter coupled system was previously experimentally demonstrated and theoretically considered in other systems \citep{Coles2011,Somaschi2011,Michetti2009,Mazza2013}. Additionally, the minimal observed lasing threshold in the weakly coupled regime ($M=17$ in Fig.\ref{Fig4}) exceeds the lowest polariton lasing threshold at the optimal detuning (corresponding to $M=3$ and $M=5$ in Fig.\ref{Fig4}) by approximately a factor of eighteen, while at suboptimal detunings the weak-to-strong coupling threshold ratio exceeds one order of magnitude.

\section{Discussion}

The experimental results demonstrate the decisive role of cavity length and coupling strength in governing the lasing dynamics of MeLPPP-based organic microcavities.
The observed increase in lasing threshold (approximately eighteen-fold when comparing optimal conditions) across the strong-to-weak coupling transition unambiguously distinguishes polariton lasing, driven by bosonic stimulation, from conventional photon lasing that relies on population inversion. While a threshold contrast of comparable magnitude between the polariton and photon lasing regimes has been reported previously in laminated fluorescent-protein microcavities \cite{Dietrich2016}, where authors observe both thresholds at each detuning position as sequential pump-power transitions within the strong-coupling regime using a wedged cavity that scans detuning on a single sample. Here, by continuously tuning the effective cavity length within a single MeLPPP device, we resolve the strong-to-weak coupling transition as a continuous function of detuning rather than as two endpoint observations, allowing Rabi splitting, effective mass, and lasing threshold to be tracked along the same specimen from $M=3$ to $M=17$.

Beyond the threshold comparison itself, the quantitative decomposition of the nonlinear energy shifts (Fig.\ref{Fig3}) provides a second independent signature of the coupling regime. The V-shaped dependence of the blueshift on spectral offset from the ASE peak is universal across all three regimes, confirming the conclusion of Ref.~\citep{Yagafarov2020} that blueshifts alone do not distinguish polariton from photon lasing. The baseline $E_0$ extracted from the parabolic fit is within the fit uncertainty consistent with zero in the weakly coupled regime and marginally positive ($E_0 = +0.3$ meV) at $M=5$, in qualitative agreement with the Rabi-quenching prediction of Ref.~\citep{Yagafarov2020} but not at a statistical confidence level that would support it as a stand-alone diagnostic. A definitive demonstration of the baseline as a coupling-regime fingerprint requires data at deeper strong coupling and with improved per-point precision, which we identify as a direction for future work.

In the strong coupling regime, the lasing emission exhibits characteristic features of polariton condensation, including a nonlinear increase of the polariton PL intensity, spectral blueshift, and linewidth narrowing while crossing the condensation threshold. As the system transitions into the weak coupling regime, these signatures diminish, and the emission energy converges towards the material gain maximum, reflecting the onset of photon lasing behavior. The ability to systematically tune the effective cavity length thus provides a powerful means of probing the continuous evolution between hybrid light–matter and purely photonic lasing regimes within a single material platform. Moreover, the correlation between reduced lasing threshold and vibronic resonance conditions highlights the important role of vibron-assisted exciton relaxation in enabling efficient polariton formation and condensation in conjugated polymers, consistent with theoretical predictions \cite{Michetti2009,Mazza2013} and prior experimental observations in organic microcavities \cite{Zasedatelev2019,Coles2011,Somaschi2011}. Our systematic study across multiple cavity lengths and mode orders provides additional evidence for this mechanism and extends its characterisation into the intermediate and weak-coupling regimes, where the threshold modulation at the vibronic resonances smoothly weakens as the coupling strength is reduced.

These findings reinforce MeLPPP as a highly suitable model system for exploring fundamental aspects of strong coupling and polariton dynamics in organic semiconductors. Notably, the asymmetric open-cavity geometry employed here, with the active polymer layer on one side of the air gap, represents one possible configuration. A symmetric design with polymer layers on both cavity mirrors could potentially increase the effective oscillator strength and extend the strong-coupling regime to larger detunings. However, such a configuration would also increase absorption losses, and the net effect on lasing threshold would depend on the balance between enhanced coupling and increased optical losses. Optimization of cavity symmetry and field distribution represents an interesting direction for future device engineering. Looking forward, engineering of cavity architectures and photonic lattice configurations could enable control of vibronic interactions and polariton dispersion, paving the way toward low-threshold, room-temperature polariton devices and integrated organic polaritonic circuits.

\section{Methods}
\textbf{Sample fabrication.} The DBRs were fabricated by sputter-depositing of 12.5 pairs of alternating \ce{Ta2O5}/\ce{SiO2} quarter wavelength-thick layers onto a sapphire substrate with \ce{Ta2O5} terminating layer. 

To prepare the precursor solution at a concentration of 1 mM, 21 mg of MeLPPP was dissolved in 1 mL of toluene via sonication for 5 minutes. A 1 cm$^2$ substrate featuring a DBR was cleaned sequentially in ultrasonic bath using acetone, isopropanol and deionized water, with 1 minute for each step then dried using nitrogen. Subsequently, the substrate underwent treatment in a plasma asher for 15 minutes in a mixture of argon and nitrogen to enhance surface hydrophilicity. 

For the deposition of MeLPPP, the precursor solution was spin-coated onto the substrate at a speed of 5000 rpm for 1 minute, with a ramp rate of 500 rpm/s. During this process, 30 µL of the precursor solution was dynamically dispensed once the desired rotational speed was achieved.
Finally, a second DBR (on a separate sapphire substrate) was positioned above the MeLPPP-coated surface using a point-squeeze mounting apparatus. The apparatus applies localized pressure at a single point, creating an open-cavity geometry where the air gap between the polymer surface and the top DBR increases radially from the contact point. This configuration allows continuous tuning of the effective cavity length by translating the sample laterally. No adhesive bonding or thermal treatment is used; the cavity assembly is maintained by the mechanical mounting.

An additional non-cavity reference MeLPPP film was spin-coated onto \ce{SiO2}/\ce{Si} substrate under conditions completely identical to the cavity’s active layer.

The active layer thickness of 180$\pm$20 nm in the cavity was estimated from profilometer (Alpha-Step D-600, KLA-Tencor (USA)) measurements of a reference non-cavity MeLPPP film.\\
\\
\textbf{Spectroscopy.} The absorption measurements of the MeLPPP reference non-cavity film were performed using a Lambda 1050 UV/Vis/NIR spectrometer (PerkinElmer).

ASE measurements of the reference film were performed by using a high energy Ti:Sapphire regenerative amplifier (Coherent Libra-HE) providing $\sim$250 fs pulses at a 500 Hz repetition rate. The amplifier's output was then converted to 400 nm via second-harmonic generation in a BBO crystal. A 50 mm cylindrical lens was used to focus the beam on the sample creating a stripe excitation profile (1790 $\mu$m x 95 $\mu$m). Stimulated emission of a reference non-cavity MeLPPP film was detected from the edge of the film, in the direction of the stripe and perpendicular to the propagation direction of the incident pump beam using Ocean Optics QE PRO spectrometer.

For real- and momentum-space imaging of the MeLPPP microcavity output emission, excitation optical pulses of $\approx$250 fs duration at 400 nm, generated by frequency-doubling the output of a 500 Hz high energy Ti:Sapphire regenerative amplifier (Coherent Libra-HE) in BBO crystal, were used. An optical beam was spectrally filtered using a long-pass filter Semrock FF01-750/LP-25 and focused onto the microcavity with a Nikon Plan Apo 20X microscopic objective, producing a Gaussian spot with a diameter of $\approx$10 $\mu$m at the $1/e^{2}$ intensity level. Photoluminescence was collected in the transmission configuration (a schematic of the experimental setup is provided in Section 4 in SI). After passing through the collection objective (Mitutoyo Plan Apo 50X), which provides a $\approx\pm$36$^{\circ}$ detection window in reciprocal space, the emission was spectrally filtered (ThorLabs LP442) and coupled into the spectrometer (Princeton Instruments ProEM 1024BX). The Fourier plane of the collection objective was projected onto the entrance slit of the spectrometer using an additional conjugate lens. An additional imaging camera (Hamamatsu Orca) was configured for the convenient control of the excitation beam focus adjustment and cavity thickness tuning. White light reflectivity measurements were performed using an Ocean Optics DH-2000 fiber-coupled Halogen–Deuterium white light source.\\
\\
\textbf{Threshold determination.} The lasing threshold was determined by measuring the integrated emission intensity within a narrow angular range of \(k_{||}=0\pm5^\circ\) as a function of the pump fluence. The data were plotted on a double logarithmic scale, and linear fits were applied separately to the spontaneous emission regime (below threshold) and the stimulated emission regime (above threshold). The threshold fluence value was defined as the intersection point of these two linear approximations. Each cavity configuration was measured 3--5 times, with the final threshold fluence value derived from the combined set of measurements.
\\
\\
\section{Author statements}
\textbf{Acknowledgement:}
Authors acknowledge Dr. Daria Khmelevskaia for providing electric field distribution simulation.\\
\\
\textbf{Funding information:}
This work was supported by the Russian Science Foundation (RSF) Grant No. 25-72-00149.\\
\\
\textbf{Author contributions:}
All authors have accepted responsibility for the entire content of this manuscript and consented to its submission to the journal, reviewed all the results and approved the final version of the manuscript.\\
\\
\textbf{Conflict of interest:}
The authors declare no conflict of interest.\\
\\
\textbf{Informed consent:}
Informed consent was obtained from all individuals included in this study.\\
\\
\textbf{Data availability statement:}
The data that support the findings of this study are available from the corresponding author upon request.\\
\\

\bibliographystyle{ieeetr}
\bibliography{Bibliography}

\setcounter{equation}{0}
\setcounter{figure}{0}
\setcounter{section}{0}
\setcounter{subsection}{0}
\newcolumntype{P}[1]{>{\centering\arraybackslash}p{#1}}
\newcolumntype{M}[1]{>{\centering\arraybackslash}m{#1}}
\renewcommand{\theequation}{S\arabic{equation}}
\renewcommand{\thefigure}{S\arabic{figure}}
\renewcommand{\baselinestretch}{1}
\onecolumngrid
\newpage
\vspace{1cm}
\begin{center}
\Large \textbf{Supplementary Information}
\end{center}

\section{Synthesis procedure of MeLPPP precipitate}

Monomer 1 and monomer 2 were synthesized according to published procedures \citep{Ammenhauser2020,Zhai2021,Bonn2016}. Polymer 1 was synthesized using an adapted methodology established for ladderane-type polymers \citep{Scherf1991}. For this purpose, a fresh Pd(PPh$_3$)$_4$ catalyst was prepared \citep{noauthor1984} and used the following day. Polymer 1 was functionalized with a 100-fold excess of MeLi relative to the monomeric unit equivalent \citep{Scherf1992}. A fresh MeLi solution was also prepared according to an adapted procedure \citep{noauthor1984} and used on the same day. The $^1$H NMR spectrum registered in d8-toluene is in full agreement with the literature data, showing the appearance of characteristic signals for OH and Me groups in a 1:3 ratio in the 2-3 ppm region compared to the spectrum of the initial Polymer 1 \citep{Scherf1992}. The final synthesis of Me-LPPP was performed according to a literature procedure \citep{Scherf1992}. To remove oligomers, the product was dissolved in a minimal amount of toluene, and acetone was added until precipitate flocculation was observed. After allowing the mixture to stand for 2-3 hours, the polymer was separated from the mother liquor by centrifugation and decantation. The resulting Me-LPPP precipitate was washed three times with acetone and dried on a vacuum line.

\begin{figure*}[h!]
    \centering
    \includegraphics[width=0.8\textwidth]{FigureS1.pdf}
    \caption{\justifying Step-by-step chemical process of MeLPPP polymer synthesis.}
    \label{fig:Chem}
\end{figure*}

\newpage

\section{Amplified Spontaneous Emission (ASE) measurements} 

Amplified spontaneous emission behavior was characterized using the experimental setup shown in Fig.\ref{fig:ASE}. The polymer layer forms a planar waveguide, as its refractive index at the emission wavelength of 400 nm exceeds that of both the \ce{SiO2}/Si substrate and the surrounding air. Upon increasing the excitation fluence above a threshold of $\approx$ 4.9 $\mu$J cm$^{-2}$, the photoluminescence (PL) spectrum collapsed into a narrow peak centered at 487.25 nm, indicating the onset of ASE.

\begin{figure*}[h!]
    \centering
    \includegraphics[width=0.9\textwidth]{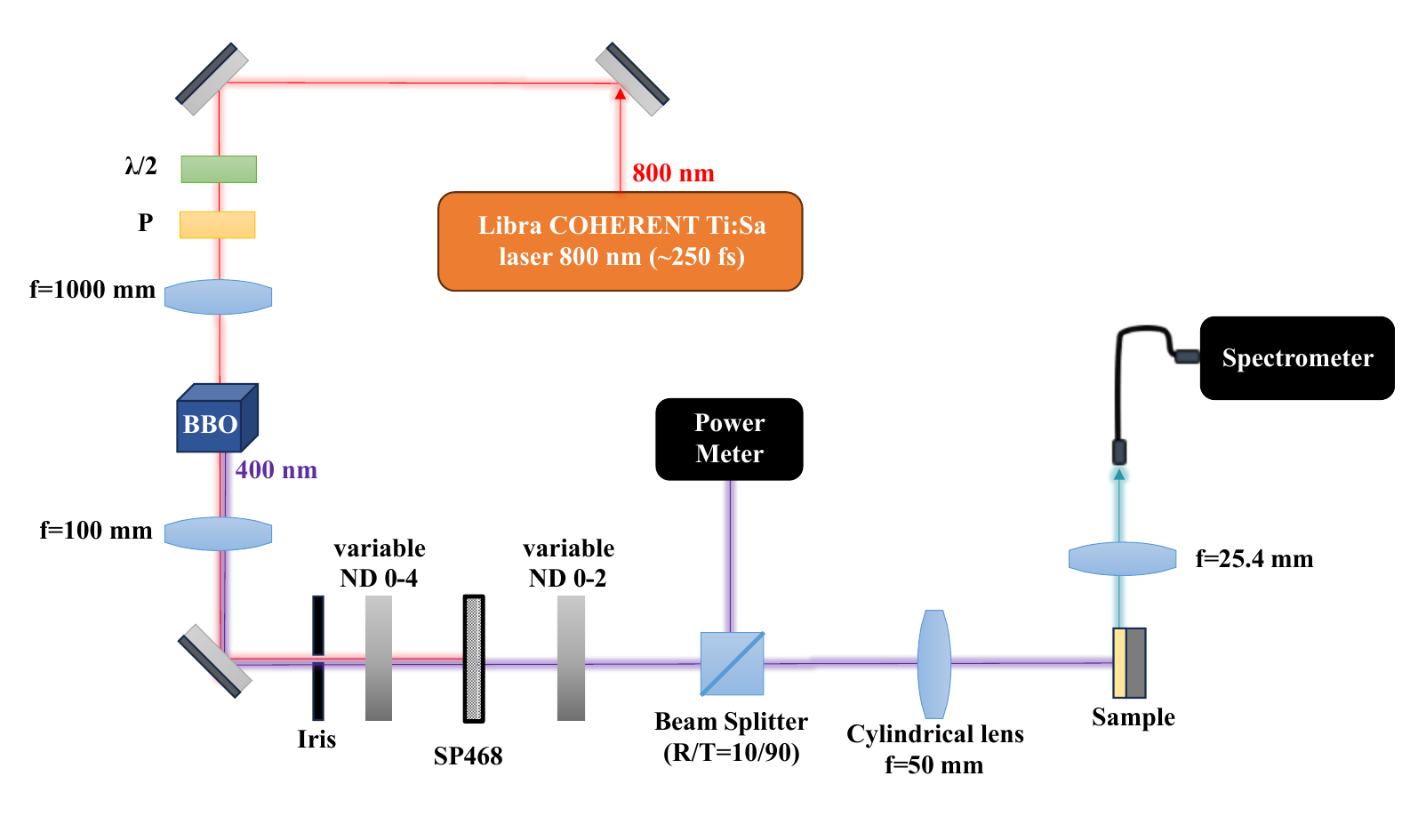}
    \caption{\justifying Schematic of the experimental setup for optical characterisation of amplified spontaneous emission of a MeLPPP thin film. Optical elements designations are provided in Fig.\ref{fig:PL}}
    \label{fig:ASE}
\end{figure*}

\newpage

\section{Optical properties of the MeLPPP-containing DBR microcavity}
\label{sec:numeric}
Numerical calculations of the reflectance spectra and electric-field distribution were performed in COMSOL Multiphysics using the finite element method. A 2D model with periodic boundary conditions (Floquet periodicity) and two periodic input/output ports was employed. The microcavity consisted of bottom and top distributed Bragg reflectors (12.5 pairs of Ta$_2$O$_5$/SiO$_2$ $\lambda$/4 layers) with a MeLPPP polymer layer in between. The refractive indices were taken as n = 2.06 for the 56-nm Ta$_2$O$_5$ layer, n = 1.487 for the 85-nm SiO$_2$ layer, and n = 1.8 for the 180-nm MeLPPP layer.

\begin{figure*}[h]
    \centering
    \includegraphics[width=0.8\textwidth]{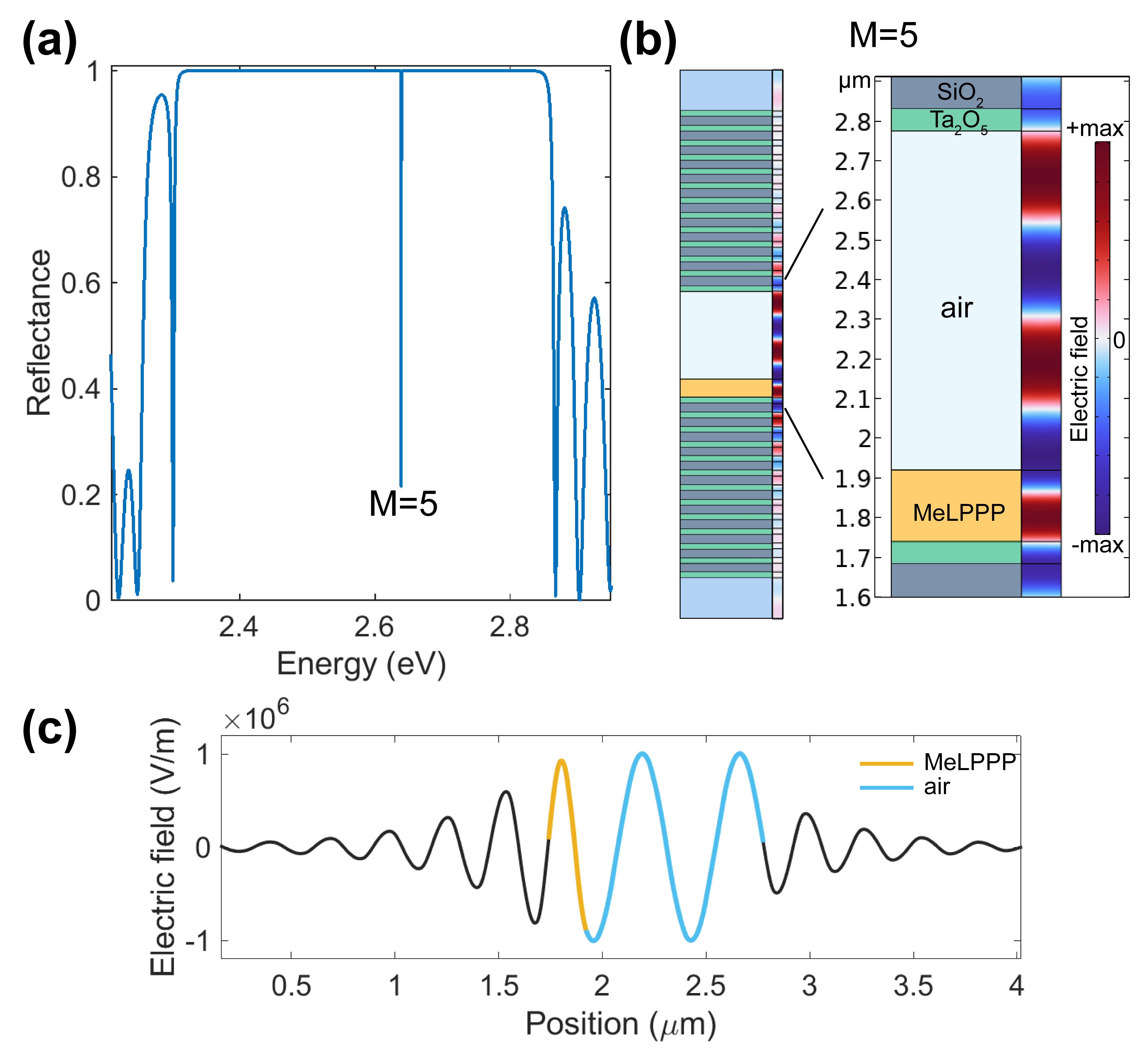}
    \caption{\justifying (a) Calculated reflectance spectrum of a microcavity with cavity mode order M = 5. (b) Schematic of the microcavity structure and spatial distribution of the out-of-plane electric field at the resonance of the M = 5 cavity mode (left panel). Magnified view of the microcavity region (right panel). (c) Spatial profile of the electric-field amplitude along the structure.}
    \label{fig:fields}
\end{figure*}

Figure \ref{fig:fields}a shows the reflectance spectrum calculated for a microcavity with an 855-nm air gap between the MeLPPP layer and the top DBR. The pronounced mode within the stopband corresponds to the cavity mode of order M = 5, as confirmed by the electric-field distribution (Figure \ref{fig:fields}b). The out-of-plane electric-field profile shows that one of the field antinodes is located within the polymer layer (Figure \ref{fig:fields}c).

\newpage

\section{Derivation of the V-shape energy-shift model}
\label{sec:vshape}

The main manuscript (Eq.~2) introduces a quantitative framework that links the spectral dependence of the lasing threshold to the spectral dependence of the nonlinear energy shift via a parabolic, saturation-driven V-shape. In this section we derive that expression from the saturation picture of Yagafarov \textit{et al.}~\citep{Yagafarov2020}, identify each of the three contributions, and quantify their relative magnitudes.

\subsection{Physical mechanisms}
\label{subsec:mechanisms}

We have conducted an extensive series of measurements with careful attention to alignment and reproducibility, and are now confident in the observed energy-shift dependence on ASE offset across all cavity mode orders studied. The total nonlinear energy shift arises from three mechanisms, which we summarise here before assembling them into the V-shape model.

\paragraph{Saturation-induced blueshift.}
Non-resonant optical pumping excites intracavity molecules, saturating their optical transitions via Pauli blocking. This produces two effects:
\begin{enumerate}
    \item quenching of the vacuum Rabi splitting for strongly coupled molecules (Eq.~2 of Ref.~\citep{Yagafarov2020});
    \item renormalisation of the cavity mode energy via the Kramers--Kronig relation for all molecules, including the weakly coupled majority.
\end{enumerate}
Both effects scale linearly with the saturation parameter
\begin{equation}
    \xi = \frac{n_x}{n_0},
\end{equation}
where $n_x$ is the density of excited molecules and $n_0$ is the total molecular density. In the small-saturation limit ($\xi \ll 1$), the total blueshift from these mechanisms is given by Eq.~8 of Ref.~\citep{Yagafarov2020}:
\begin{equation}
    \Delta E_{\rm LPB} = \frac{\xi}{2}\left[
        \frac{s\,\hbar\Omega_0}{\sqrt{1+s^2}}
        + \frac{(E_x - |\delta|)\,F[d]\,\alpha}{5\,n_{\rm eff}}\left(1 + \frac{1}{\sqrt{1+s^2}}\right)
    \right],
    \label{eq:yagafarov}
\end{equation}
where $s = \hbar\Omega_0/|\delta|$ is the strong-coupling parameter, $F[d]$ is the Dawson function encoding the Kramers--Kronig spectral dependence, and $\alpha$ is the oscillator-strength parameter. The critical point is that both terms are linear in $\xi$, so the blueshift magnitude is directly proportional to the saturation parameter.

\paragraph{Gain-induced frequency pulling.}
The lasing mode is pulled toward the gain maximum (the ASE peak) by an amount proportional to the ratio of the cavity linewidth to the gain linewidth. This produces an antisymmetric contribution: a redshift when the cavity mode is on the blue side of the ASE peak and a blueshift on the red side. For our system, the cavity linewidths ($\Delta\omega_c \sim 1.5$--$10$~meV) are much narrower than the gain bandwidth ($\Delta\omega_{\rm gain} \sim 40$~meV), making this contribution small.

\paragraph{Threshold-dependent saturation parameter.}
The key insight connecting these mechanisms to the observed spectral dependence of the energy shifts is that the saturation parameter $\xi$ \emph{at threshold} is not constant across the gain spectrum, but varies systematically with the spectral offset $\Delta E = E_{\rm ASE} - E_{k_0}$ from the ASE peak. This variation is governed by the lasing threshold, which we have independently measured for each data point.

\subsection{Derivation of the quadratic model}
\label{subsec:derivation}

We now derive the functional form of the energy-shift dependence on spectral offset. The derivation proceeds in three steps.

\subsubsection*{Step 1: Parabolic threshold dependence}

At the lasing threshold, the round-trip gain must equal the round-trip losses. The net gain $g(E)$ available at cavity mode energy $E$ is determined by the spectral overlap between the cavity resonance and the material gain profile, with the lasing threshold reached when the round-trip gain compensates the round-trip losses. The gain profile of MeLPPP, as characterised by the ASE spectrum of the bare film, is a smooth, single-peaked function of energy centred at $E_{\rm ASE}$. By definition, the first derivative of any smooth function vanishes at its maximum, so the leading-order variation near $E_{\rm ASE}$ is quadratic --- a direct consequence of the Taylor expansion to second order:
\begin{align}
    g(E) &= g(E_{\rm ASE}) + \left.\frac{\partial g}{\partial E}\right|_{E_{\rm ASE}}(E - E_{\rm ASE}) + \frac{1}{2}\left.\frac{\partial^2 g}{\partial E^2}\right|_{E_{\rm ASE}}(E - E_{\rm ASE})^2 + \ldots\nonumber\\
    &\approx g_{\max} - \tfrac{1}{2}\,|g''|\,(E - E_{\rm ASE})^2,
\end{align}
where $g''$ is the curvature of the gain spectrum at its maximum. Since the threshold pump fluence is inversely proportional to the net gain, $P_{\rm th} \propto 1/g(E)$, we obtain:
\begin{equation}
    P_{\rm th}(\Delta E) \approx P_{\rm th,\,min}\left[1 + \left(\frac{\Delta E}{w}\right)^2\right],
    \label{eq:pthparabolic}
\end{equation}
where $w$ is a characteristic gain width related to the curvature of the gain spectrum, and $P_{\rm th,\,min}$ is the minimum threshold at the ASE centre. This parabolic threshold dependence is directly confirmed by our independent threshold measurements (Fig.\ref{Fig2}d of the main manuscript, left axis), which show that the lasing threshold is minimised when the spectral offset from the exciton matches the energies of the dominant vibrational modes V$_2$ and V$_3$ ($\sim$193--198~meV, identified from the Raman spectrum of MeLPPP), corresponding to the condition of most efficient vibron-mediated relaxation from the exciton reservoir to the lasing mode.

\subsubsection*{Step 2: Saturation parameter at threshold}

Under non-resonant pumping at 400~nm, all intracavity molecules are excited with equal probability regardless of their coupling to the cavity mode. The fraction of excited molecules at threshold is:
\begin{equation}
    \xi = \frac{\sigma_{\rm abs}\,P_{\rm th}}{\hbar\omega_{\rm pump}},
\end{equation}
where $\sigma_{\rm abs}$ is the absorption cross section at the pump wavelength. Since $\sigma_{\rm abs}$ and $\hbar\omega_{\rm pump}$ are constant across all measurements (same material, same pump wavelength), we have $\xi \propto P_{\rm th}$. Substituting the parabolic threshold from Eq.~\eqref{eq:pthparabolic}:
\begin{equation}
    \xi(\Delta E) = \xi_{\min}\left[1 + \left(\frac{\Delta E}{w}\right)^2\right].
\end{equation}

\subsubsection*{Step 3: Combining contributions}

From Eq.~\eqref{eq:yagafarov}, the saturation-induced blueshift is linear in $\xi$ in the small-saturation regime ($\xi \ll 1$). Denoting the proportionality constant (which depends on $\hbar\Omega_0$, $\delta$, $n_{\rm eff}$, $\alpha$, and the FWHM) as $C$, we have:
\begin{equation}
    \Delta E_{\rm sat} = C \cdot \xi(\Delta E) = C \cdot \xi_{\min}\left[1 + \left(\frac{\Delta E}{w}\right)^2\right].
\end{equation}
Combining all contributions, the total energy shift is:
\begin{equation}
    \boxed{E_{\rm lasing} - E_{k_0} = A_{\rm sat}\left(\frac{\Delta E}{w}\right)^2 + A_{\rm pull}\,\Delta E + E_0,}
    \label{eq:vshape}
\end{equation}
where $A_{\rm sat} = C \cdot \xi_{\min}$ is the amplitude of the threshold-dependent symmetric blueshift, $A_{\rm pull}$ is the gain-pulling coefficient, and $E_0 = C \cdot \xi_{\min} + \Delta E_{\rm pull}(0)$ is the baseline blueshift at the ASE centre. Each parameter is directly traceable to a physical mechanism identified in Ref.~\citep{Yagafarov2020}.

We note that the parabolic approximation is strictly valid near the gain maximum; at large spectral offsets ($|\Delta E| \gg w$), the gain profile falls off faster than a parabola (Gaussian or Lorentzian tails), leading to a threshold that increases more steeply and correspondingly larger blueshifts than the parabolic model predicts.

\begin{figure*}[h!]
    \centering
    \includegraphics[width=0.9\textwidth]{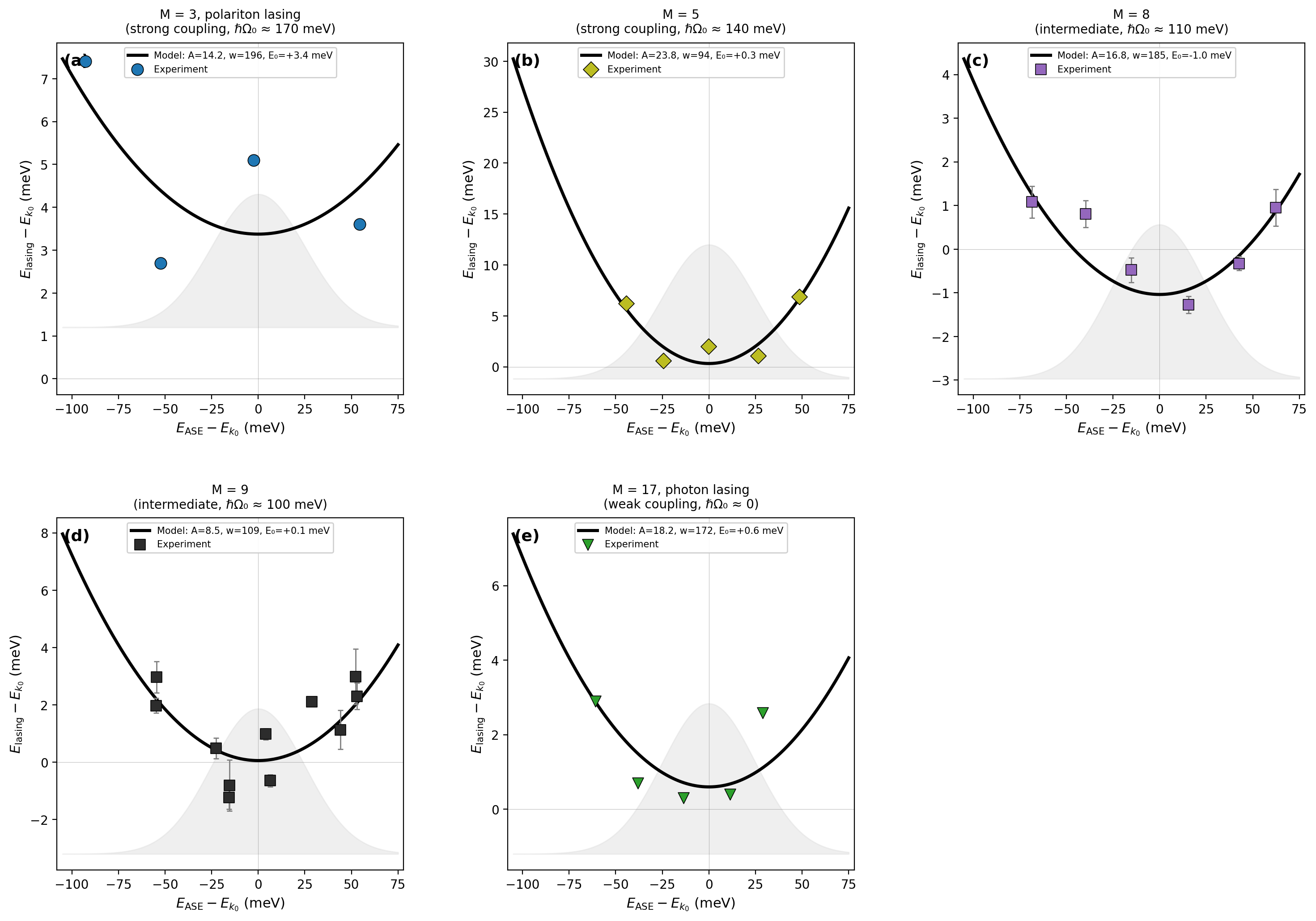}
    \caption{\justifying Parabolic V-shape fit of the nonlinear energy shift across five cavity-mode orders.
The nonlinear energy shift at $\approx 2\times P_{\rm th}$ relative to the below-threshold emission at $k_{||}=0$,
$E_{\rm lasing} - E_{k_0}$, is plotted as a function of the spectral offset from the ASE maximum,
$\Delta E = E_{\rm ASE} - E_{k_0}$, for
\textbf{a}, $M=3$;
\textbf{b}, $M=5$;
\textbf{c}, $M=8$;
\textbf{d}, $M=9$; and
\textbf{e}, $M=17$.
Black solid curves show best fits to the parabolic V-shape model,
$E_{\rm lasing} - E_{k_0} = A_{\rm sat}(\Delta E/w)^2 + E_0$
(antisymmetric component fixed at $A_{\rm pull}=0$; see Section~\ref{sec:vshape}).
Fitted parameters are given in each legend.
The grey shaded region shows the normalised ASE spectrum of a 180-nm MeLPPP reference film, plotted for visual reference.
Error bars, where shown, represent the standard deviation of repeated threshold measurements.}
    \label{fig:ASE}
\end{figure*}

\subsection{Quantitative analysis of the fitted parameters}
\label{subsec:quantitative}

The quantitative analysis reveals that the dominant contribution to the energy shift is the symmetric, threshold-dependent saturation term ($A_{\rm sat}$), which is present for all cavity orders from the strongly coupled $M = 3$ to the weakly coupled $M = 17$, consistent with the conclusion of Ref.~\citep{Yagafarov2020} that saturation-induced blueshifts operate independently of the coupling regime. The parabolic dependence on spectral offset from the ASE peak directly mirrors the independently measured parabolic dependence of the lasing threshold, which we attribute to vibron-mediated relaxation efficiency (V$_2$ and V$_3$ modes at $\sim$193--198~meV, as identified from the Raman spectrum of MeLPPP), establishing a quantitative, causal link between the threshold variation and the energy shift pattern via the saturation parameter $\xi$.

Gain-induced frequency pulling, by contrast, is negligible: the antisymmetric pulling coefficient $|A_{\rm pull}| < 0.015$ for all cavity orders, contributing less than 1~meV even at 50~meV spectral offset, which is a direct consequence of the narrow cavity linewidths relative to the broad ASE gain spectrum of MeLPPP, and consistent with Ref.~\citep{Yagafarov2020}.

While the V-shape itself is thus universal, the baseline blueshift $E_0$ differs between coupling regimes: the $M = 3$ cavity ($\hbar\Omega_0 \approx 170$~meV) exhibits $E_0 = +3.1$~meV, significantly larger than the near-zero baselines observed for $M = 8$, $9$, and $17$, consistent with residual Rabi quenching even at the minimum threshold excitation density --- a channel that is absent in the weak-coupling regime. Thus, while the observation of blueshifts alone does not distinguish polariton from photon lasing, in full agreement with Ref.~\citep{Yagafarov2020}, the quantitative decomposition of the energy shift into its spectral-offset-dependent and baseline components reveals coupling-regime-dependent information.

\newpage

\section{Optical Setup for Strong- and Weak-coupling Characterization} 

All real- and momentum-space imaging experiments described in the main text were performed using the following experimental scheme. The sample was mounted on an XYZ stage at the focal plane of the focusing objective. A beam sampler before the sample was used to direct a portion of the pump beam to a Thorlabs DET 10A photodetector connected to an oscilloscope (Picoscope 2000). The pump pulse energy was controlled using a gradient ring ND filter. Data acquisition was automated via a custom LabView (National Instruments) program, operating in a time-integrated mode below the lasing threshold and a single-shot mode above the lasing threshold. The single-shot regime allowed for the synchronous acquisition of individual pump pulse amplitudes and their corresponding photoluminescence emissions, thus preventing averaging over the intensity fluctuations of the laser. For measuring reflectivity spectrum of the microcavity an Ocean Optics DH-2000 fiber-coupled Halogen–Deuterium white light source was used.

\begin{figure*}[h]
    \centering
    \includegraphics[width=1\textwidth]{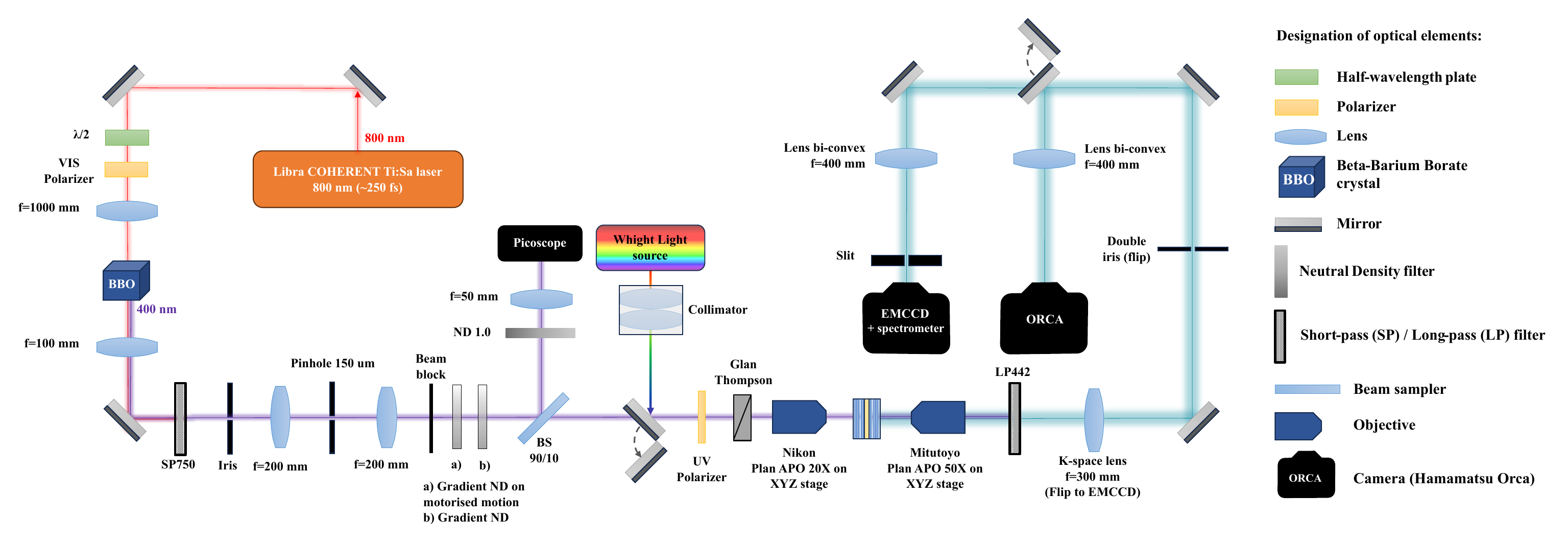}
    \caption{\justifying Schematic of the experimental setup for optical characterisation of photoluminescence emission from the investigated microcavity.}
    \label{fig:PL}
\end{figure*}

\end{document}